\documentclass[
 aps,
 prl,
 twocolumn,
superscriptaddress,
preprintnumbers,
bibnotes,
 amsmath,amssymb,
]{revtex4-2}

\usepackage{graphicx}
\usepackage{dcolumn}
\usepackage{bm}
\usepackage{natbib}
\usepackage{hyperref}

\usepackage{xcolor}

\renewcommand{\d}{\ensuremath{\mathrm{d}}}
\newcommand{\dc}{\ensuremath{\mathrm{dc}}}
\newcommand{\pbh}{\ensuremath{\mathrm{PBH}}}
\newcommand{\gw}{\ensuremath{\mathrm{GW}}}

\newcommand{\eqsp}{\ensuremath{\;}}

\begin{document}

\preprint{DESY-23-093, MITP-23-036, \href{https://doi.org/10.1103/PhysRevResearch.7.013196}{10.1103/PhysRevResearch.7.013196}}

\title{Signals of merging supermassive primordial black holes in pulsar timing arrays}

\newcommand{\AddrMPIK}{%
Max-Planck-Institut f\"ur Kernphysik, Saupfercheckweg 1, 69117 Heidelberg, Germany}
\newcommand{\AddrDESY}{%
Deutsches Elektronen-Synchrotron DESY, Notkestr.~85, 22607 Hamburg, Germany}
\newcommand{\AddrMainz}{%
PRISMA$^+$ Cluster of Excellence and Mainz Institute for Theoretical Physics, Johannes Gutenberg-Universit\"at Mainz, 55099 Mainz, Germany}

\author{Paul Frederik Depta}
\email{frederik.depta@mpi-hd.mpg.de}
\affiliation{\AddrMPIK}

\author{Kai Schmidt-Hoberg}
\email{kai.schmidt-hoberg@desy.de}
\affiliation{\AddrDESY}

\author{Pedro Schwaller}
\email{pedro.schwaller@uni-mainz.de}
\affiliation{\AddrMainz}

\author{Carlo Tasillo}
\email{carlo.tasillo@desy.de}
\affiliation{\AddrDESY}

\date{July 25, 2023}

\begin{abstract}
In this work we evaluate whether the gravitational wave background recently observed by a number of different pulsar timing arrays could be due to merging \emph{primordial} supermassive black hole binaries. We find that for homogeneously distributed primordial black holes this possibility is inconsistent with strong cosmological and astrophysical constraints on their total abundance. If the distribution exhibits some clustering, however, the merger rate will in general be enhanced, opening the window for a consistent interpretation of the pulsar timing array data in terms of merging primordial black holes, if $\mu$-distortion constraints associated with the formation mechanism can be evaded.
\end{abstract}

\maketitle


\paragraph*{Introduction.---}%
The discovery of gravitational waves (GWs) with frequencies $\mathcal{O}(100)\,\mathrm{Hz}$ from binary black hole mergers in 2015~\cite{LIGOScientific:2016aoc} has opened new possibilities for the exploration of our Universe. Depending on the astrophysical or cosmological source, GW signals may exist over a wide range of frequencies and complementary experimental approaches aim to explore much of the available parameter space~\cite{LISA:2017pwj, NANOGrav:2020bcs}. Pulsar timing arrays (PTAs) in particular are sensitive at frequencies in the nHz range. In 2020, the North American Nanohertz Observatory for Gravitational Waves collaboration (NANOGrav)~\cite{NANOGrav:2020bcs} announced evidence for a common red process hinting at a stochastic nHz gravitational wave background (GWB), with similar results subsequently published by other PTAs~\cite{Goncharov2021,Chen2021,Antoniadis2022}.
Excitingly, the very recent findings from NANOGrav~\cite{NANOGrav:2023gor}, the European PTA (EPTA)~\cite{Antoniadis:2023ott}, Parkes PTA (PPTA)~\cite{Reardon:2023gzh}, and Chinese PTA (CPTA)~\cite{Xu:2023wog} not only strengthen this signal, but also show mounting evidence for the characteristic quadrupolar 
Hellings-Downs correlation~\cite{Hellings1983}. It is, therefore, fascinating and timely to investigate possible origins of this newly observed signal in terms of a gravitational wave background (GWB). 
Mergers of supermassive black hole binaries (SMBHB) are expected to contribute to the GWB in this frequency range~\cite{NANOGrav:2020bcs}, although the local SMBHB density would need to be $\mathcal{O}(10)$ larger than previously 
estimated~\cite{Casey-Clyde2021,Kelley2016,Kelley2017} to match the observed signal amplitude.
Hence, the validity of this explanation is under active debate~\cite{Middleton2021,Izquierdo-Villalba2021,Curylo2021,Somalwar2023} and it is interesting to consider alternative sources.
A nHz GWB could also originate from truly cosmological sources~\cite{NANOGrav:2023hvm,Antoniadis:2023zhi}, such as 
inflation~\cite{Vagnozzi:2020gtf}, topological defects~\cite{Blasi2020,Ellis2020b, Buchmuller:2020lbh,Samanta:2020cdk,Guo:2023hyp,King:2023cgv}, first-order phase 
transitions~\cite{Nakai2020,Ratzinger:2020koh,NANOGrav:2021flc,Bringmann:2023opz,Madge:2023cak},
or the \emph{production} of sub-solar mass primordial black holes (PBHs)~\cite{DeLuca:2020agl,Vaskonen:2020lbd,Kohri:2020qqd,Unal:2020mts,Ashoorioon:2022raz}.

In this work we study the possibility that the observed GW signal could be due to inspiraling supermassive PBH binaries. 
To obtain the correct signal amplitude, a sizable abundance of PBHs is required, which is subject to strong constraints from various observational probes~\cite{Carr:2020gox}. It is, therefore, a non-trivial question as to whether inspiraling PBHs could constitute a viable explanation. Indeed, we find that it is \emph{not} possible to consistently explain the NANOGrav data with homogeneously distributed PBHs. Specifically, the parameter regions at very large PBH masses which naively allow to fit the PTA data (as done in~\cite{Atal:2020yic}) no longer result in a stochastic (or even continuous) signal. For clustered PBHs, on the other hand, the formation of binary systems is more efficient and the merger rate is generally enhanced, opening the possibility for an overall consistent description of the observed GWB.

This work is structured as follows. We first review and extend the formalism required to evaluate the GWs from binary mergers for the case of clustered PBHs. After discussing the relevant constraints and how to match the predicted signal to the PTA data we present our results.

\begin{figure*}[t]
    \centering
    \includegraphics[width=\columnwidth]{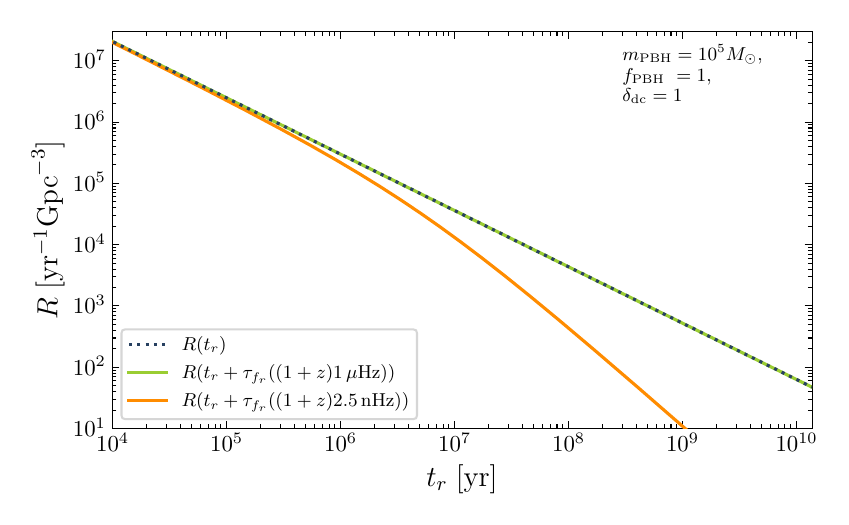}
    \includegraphics[width=\columnwidth]{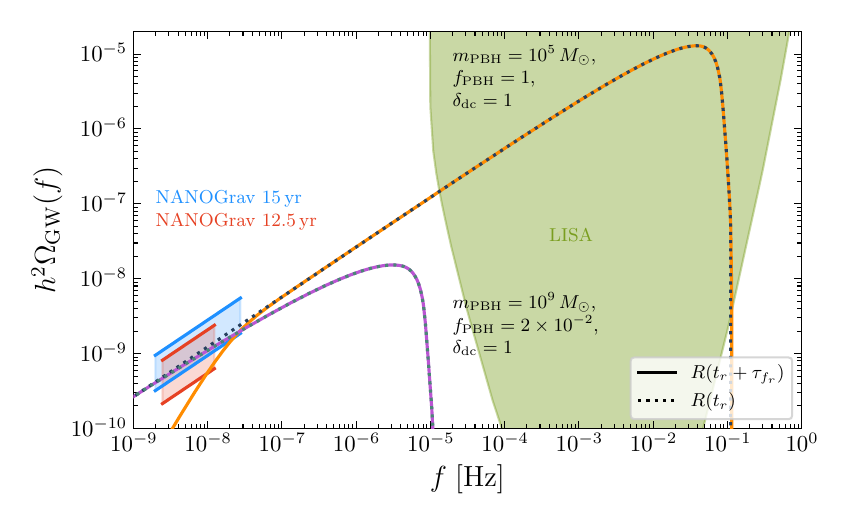}
    \caption{Left: Rate for binaries merging at $t_\text{r}$ (dotted line) as well as rate for emitting GWs at $t_\text{r}$ with different frequencies today (solid green and orange lines). Right: GW energy density $h^2\Omega_\text{GW}(f)$ obtained using the correct rate $R(t_\text{r} + \tau_{f_\text{r}})$ (solid lines) and instead using the merger rate $R(t_\text{r})$ in Eq.~\eqref{eq:omega_gw} (dotted lines). We also show the regions of the NANOGrav 12.5\,yr and 15\,yr signals assuming the inspiral of binary BHs. The LISA sensitivity curve is shown in green.}
    \label{fig:rate_ogw}
\end{figure*}

\paragraph*{Gravitational wave signal.---}%
A stochastic GWB can be specified in terms of the energy density parameter $h^2 \Omega_\gw (f)$ of the gravitational waves (GWs) per logarithmic frequency interval around a frequency $f$. In our scenario, this GWB is generated from PBH binary mergers occurring at a rate $R (t_\text{r})$ per comoving volume $V_\text{c}$ and cosmic time $t_\text{r}$~\footnote{Cosmic time refers to the time in the corresponding cosmic rest-frame, related to time $t$ measured in the present cosmic rest-frame by $\d t = (1+z) \, \d t_\text{r}$.}. The frequencies accessible to PTAs can be well below the maximal frequency emitted during the inspiral and might therefore have been radiated long before coalescence, which we need to take into account. In particular, a given binary merging at time $t_{r, \text{merg}}$ emits a frequency $f_\text{r}$ in the cosmic rest frame, redshifted to a frequency $f = f_\text{r} / (1+z)$ today, at time $t_\text{r} = t_{r, \text{merg}} - \tau_{f_\text{r}}$, where
\begin{align}
    \tau_{f_\text{r}} = \frac{5 \times  2^{1/3}}{256 \, \pi^{8/3}} \, f_\text{r}^{-8/3} \, (G \, m_\pbh)^{-5/3} \label{eq:tau_f_r}
\end{align}
is the time until coalescence for $f_\text{r}$~\cite{Maggiore:2007ulw}. As $t_{r, \text{merg}}$ is the argument for the \textit{merger} rate, the number of events per comoving volume and cosmic time emitting a frequency $f_\text{r}$ in the cosmic rest frame at cosmic time $t_\text{r}$ is given by $R (t_\text{r} + \tau_{f_\text{r}})$. To the best of our knowledge this frequency-dependent rate has not been discussed yet in the literature concerning GWs generated by PBHs. The GW energy density parameter is given by (cf.~\cite{Phinney:2001di})
\begin{align}
    \Omega_\gw (f) = \frac{f}{\rho_\mathrm{crit}} \int_0^{t_0} \d t_\text{r} \left( R (t_\text{r} + \tau_{f_\text{r}}) \frac{\d E_\gw}{\d f_\text{r}} \right)_{f_\text{r} = (1+z) f} \eqsp, \label{eq:omega_gw}
\end{align}
where $t_0$ is the current cosmic time and $\d E_\gw / \d f_\text{r}$ is the GW spectrum, which we adopt from~\cite{Ajith:2007kx}. Note that $h^2 \Omega_\gw (f)$ is, in fact, a cosmological average of many PBH binaries. We discuss the effect of only having access to a local realization of the GWB within PTA measurements below.

For the merger rate we adapt the calculation from~\cite{Raidal:2017mfl}. We assume a monochromatic initial PBH mass distribution, expecting that an extended distribution does not qualitatively change our results. A PBH binary forms in the early Universe when the gravitational attraction between two neighboring PBHs overcomes the Hubble flow. A third close-by PBH provides angular momentum such that the two PBHs do not simply collide 
and typically have a large eccentricity~\cite{Nakamura:1997sm,Ioka:1998nz,Sasaki:2016jop}. After the first version of this work appeared, it was noted that the high eccentricities characteristic for our model setup actually gave a good fit to the observed GW spectral shape~\cite{Raidal:2024odr} and lowered its degree of anisotropy~\cite{Raidal:2024tui}.

Clearly the merger rate depends on the local PBH density at binary formation. It is, therefore, interesting to study the effect of clustering, 
which increases the local density by $\delta_\dc$ compared to the
global comoving number density $n_\pbh$.
The density contrast $\delta_\dc$ can be considered constant on the scales relevant for binary formation~\cite{Raidal:2017mfl}. By considering the number density of three-body configurations relevant for the PBH binary formation and their coalescence times (in terms of the distances in the three-body configurations)~\cite{Peters:1964zz,Raidal:2017mfl,Note2}\footnotetext{Using the coalescence time makes sure that our scenario does not suffer from the so-called ``final parsec problem''~\cite{Milosavljevic:2002ht} as only binaries with sufficiently short coalescence times contribute to the GWB.} one finds~\cite{Raidal:2017mfl}
\begin{align}
    &R (t_\text{r}) = \frac{9 \, \tilde{N}^{53/37}_\pbh}{296 \pi \, \delta_\dc \, \tilde{x}^3 \, \tilde{\tau}} \left( \frac{t_\text{r}}{\tilde{\tau}} \right)^{-34/37} \times \nonumber \\
    &\!\left( \Gamma \left[ \frac{58}{37}, \tilde{N}_\pbh \left( \frac{t_\text{r}}{\tilde{\tau}} \right)^{3/16} \right] \!\! - \! \Gamma \left[ \frac{58}{37}, \tilde{N}_\pbh \left( \frac{t_\text{r}}{\tilde{\tau}} \right)^{-1/7} \right] \right) \label{eq:merg_rate}
\end{align}
with $\tilde{x}^3 = 6 \,m_\pbh / (4 \pi \, a_\mathrm{eq}^3 \, \rho_\mathrm{eq})$, the scale factor $a_\mathrm{eq}$ and energy density $\rho_\mathrm{eq}$ at matter-radiation equality~\cite{Planck2018}, $\tilde{\tau} = 3 \, (a_\mathrm{eq} \, \tilde{x})^4/[170 \, (G \, m_\pbh)^3]$, $\tilde{N}_\pbh = 4 \pi \, n_\pbh \, \delta_\dc \, \tilde{x}^3 / 3$, and the incomplete gamma function $\Gamma$.

Larger clustering generally leads to a larger number of PBH binaries contributing to the merger rate (note that $\tilde{N}_\pbh \propto \delta_\dc$) as well as a shorter timescale for the mergers (cf.\ the arguments of the gamma functions). Furthermore, multiple merger steps can be possible for large clustering due to successive hierarchical merging of PBH binaries with increasing mass. When discussing the case of significant clustering we implement multiple merger steps by adding the rates and contributions to the GW energy density parameter for the corresponding steps as detailed in~\cite{Bringmann:2018mxj}. Note that this shifts our results for significant clustering to at most $\mathcal{O}(10 \, \%)$ lower PBH abundances compared to only considering a single step.

By substituting $t_\text{r} \rightarrow t_\text{r} + \tau_{f_\text{r}}$ in Eq.~\eqref{eq:merg_rate} we can compute the rate $R(t_\text{r} + \tau_{f_\text{r}})$ for the emission of a frequency $f_\text{r}$. 
To illustrate the importance of the emission time of a given frequency, in Fig.~\ref{fig:rate_ogw} (left) we compare the rate for binaries merging at $t_\text{r}$ (dotted blue line) with the rate for GW emission with frequencies $f = 1 \, \mu\mathrm{Hz}$ and $2.5 \, \mathrm{nHz}$ today (solid green and orange lines), assuming a PBH mass $m_\pbh = 10^5 \, M_\odot$, a fraction $f_\pbh = 1$ of dark matter (DM) in PBHs, and $\delta_\dc = 1$.
For instance, at $t_\text{r} = 10^8 \, \mathrm{yr}$ (i.e.\ $z \approx 30$) one obtains $\tau_{f_\text{r}}[(1+z)1 \, \mu\mathrm{Hz}] \approx 130 \, \mathrm{yr}$ and $\tau_{f_\text{r}}[(1+z)2.5 \, \mathrm{nHz}] \approx 1.1 \times 10^9 \, \mathrm{yr}$. 
Hence, the rate for the emission of GWs with the larger frequency (solid green line) is very close to the merger rate (dotted blue line), whereas the rate for the emission of GWs with the smaller frequency (solid orange line) differs significantly, i.e.\ it takes the value that the dotted blue line attains $1.1 \times 10^9 \, \mathrm{yr}$ later.

In the right panel of Fig.~\ref{fig:rate_ogw} we show $h^2 \Omega_\gw(f)$ according to Eq.~\eqref{eq:omega_gw} (solid lines) as well as just inserting the merger rate $R (t_\text{r})$ instead of $R(t_\text{r} + \tau_{f_\text{r}})$ in the integral (dotted lines). 
The difference between those calculations is especially important for the low frequencies observed by NANOGrav if the PBH mass is relatively light.

We close the discussion of the GWB signal by mentioning some assumptions in the calculation. These include a monochromatic mass distribution~\cite{Raidal:2017mfl}, the modeling of PBH clustering for multiple merger steps~\cite{Bringmann:2018mxj}, a circular orbit of the binaries for the GWB spectrum and the time until coalescence for a given frequency, neglecting the effect of other PBHs on the binaries~\cite{Ali-Haimoud:2017rtz,Raidal:2018bbj,Vaskonen:2019jpv} as well as further environmental effects e.g.\ due to interactions with surrounding inhomogeneities~\cite{Ali-Haimoud:2017rtz}, DM halos surrounding the PBH binaries~\cite{Kavanagh:2018ggo}, accretion, and late-time formation of binaries~\cite{Raidal:2017mfl}. These effects can have an $\mathcal{O}(1)$ impact on the merger rate, but often require further model-dependent assumptions, especially when clustering is involved, as well as expensive $N$-body simulations that are beyond the scope of this work. Although this may lead to a quantitative change of our results, slightly shifting the best-fit regions and constraints, our results are expected to be qualitatively robust and the conclusions to be unaffected.

\paragraph*{Expected number of binaries.---}%
The PTA signals are reported as stochastic GWBs. While there have been searches for signals of individual binaries in the data of different PTAs~\cite{IPTA:2023ero,Antoniadis:2023aac,NANOGrav:2023pdq}, no compelling evidence for these signals was found. For sufficiently large PBH masses and small abundances the expected signal will in general no longer resemble a stochastic background, as only very few binaries will contribute to the signal, causing inconsistency with observations, see~\cite{Agazie:2024jbf}.

The problem is exacerbated by the question of how well an actual distribution of merging binaries observed by a PTA (corresponding to a local GW energy density $\Omega_{\gw, \mathrm{loc}}$ on the length scales relevant for PTAs) would reproduce the global average of the GWB $h^2 \Omega_\gw (f) = h^2 \langle \Omega_{\gw, \mathrm{loc}} (f) \rangle$, i.e.\ how well the global mean is reproduced by the binaries that are in our past light cone. In particular, $h^2 \Omega_{\gw, \mathrm{loc}}(f)$ is \textit{not} deterministic given the model parameters (mass, abundance, and clustering of PBHs), but instead stochastic in itself. The statistics of $h^2 \Omega_{\gw, \mathrm{loc}}(f)$ can be evaluated using Markov chain Monte Carlo methods or moment generating functions~\cite{Ellis:2023owy}. Due to the considerable additional numerical effort we leave this for future studies, use the global average signal prediction $h^2 \Omega_\gw (f)$, and identify regions in parameter space where we expect a significant deviation from the global average.

Most importantly, the distribution of $h^2 \Omega_{\gw, \mathrm{loc}}(f)$ is influenced by the expected number of binaries $\bar{N} (f_-, f_+)$ contributing to the GWB in a frequency range between $f_-$ and $f_+$. In the Appendix we use~\cite{Sesana:2008mz} to show that
\begin{align}
    \bar{N} (f_-, f_+) = \! \int_{f_-}^{f_+} \frac{\d f}{f} \int_0^\infty \d z \, \frac{32 \pi \tau_{f_\text{r}} \, [d_\text{c}(z)]^2}{3 H(z)} R (t_\text{r}- \tau_{f_\text{r}})\eqsp,
\end{align}
where $d_\text{c}$ is the comoving distance and $H$ is the Hubble rate. As we fit to the 14 lowest frequencies in the NANOGrav 15\,yr data set we use $f_- = 1/T_\text{obs}^\text{15} = 1.98 \, \text{nHz}$ and $f_+ = 14/T_\text{obs}^\text{15} = 27.7 \, \text{nHz}$.

If $\bar{N} \gg 1$, many PBH binaries contribute to the GW signal and the local GWB is very close to the global value. This holds even though uncertainties due to some close-by binaries can be relevant for $\bar{N} \lesssim 100$ and can potentially cause spikes in the local GWB~\cite{Ellis:2023owy}. If $\bar{N} \sim \mathcal{O} (\text{few})$, i.e.\ the GW signal is composed of only a few binaries, the uncertainty in the signal prediction is considerable. It would then be more appropriate to search for individual GW events instead of a GWB~\cite{IPTA:2023ero,Antoniadis:2023aac,NANOGrav:2023pdq}. If $\bar{N} \ll 1$, even having a single PBH binary emitting GWs is unlikely and in most realisations $h^2 \Omega_{\gw, \mathrm{loc}} (f) = 0$. This is a feature of our model setup, which will soon be testable through the anisotropy of a GW background from a low number of sources~\cite{Agazie:2024jbf}.

\paragraph*{PBH production and constraints.---}%
There are many different production mechanisms for PBHs, see e.g.~\cite{Carr:1974nx,Belotsky:2018wph,Cotner:2017tir,Ferrer:2018uiu,Lewicki2023,Gouttenoire2023,Baker2021}. Highly clustered PBH distributions in particular are not expected for Gaussian primordial 
fluctuations~\cite{Ali-Haimoud:2018dau,Desjacques:2018wuu} (being subject to strong observational constraints~\cite{Mather:1993ij,Fixsen:1996nj} from Cosmic Background Explorer (COBE) Far Infrared Absolute Spectrophotometer (FIRAS)), but could, e.g., arise due to primordial non-Gaussianities~\cite{Young:2014oea,Matsubara:2019qzv} or from the collapse of domain walls~\cite{Belotsky:2018wph}. In this work we remain agnostic about the origin and spatial distribution of PBHs and concentrate on exploring the phenomenological impact of different assumptions. In~\cite{Hooper:2023nnl} it was recently shown that a population of supermassive and clustered PBHs can indeed be produced in inflationary models involving an additional curvaton field, while not violating $\mu$-distortion constraints.

Different astrophysical and cosmological observations place constraints on the abundance of heavy PBHs, which we adopt from~\cite{Carr:2020gox}. These limits assume a monochromatic mass function as well as a roughly homogeneous spatial distribution (no clustering). While we briefly comment on the expected impact of sizable clustering, a full re-evaluation of these limits would require going beyond some of the simple approximations made in the original derivations and is beyond the scope of this work. 
Also note that many of the different constraints come with different uncertainties and sometimes also with additional caveats.

\begin{figure*}
    \centering
    \includegraphics[width=\textwidth]{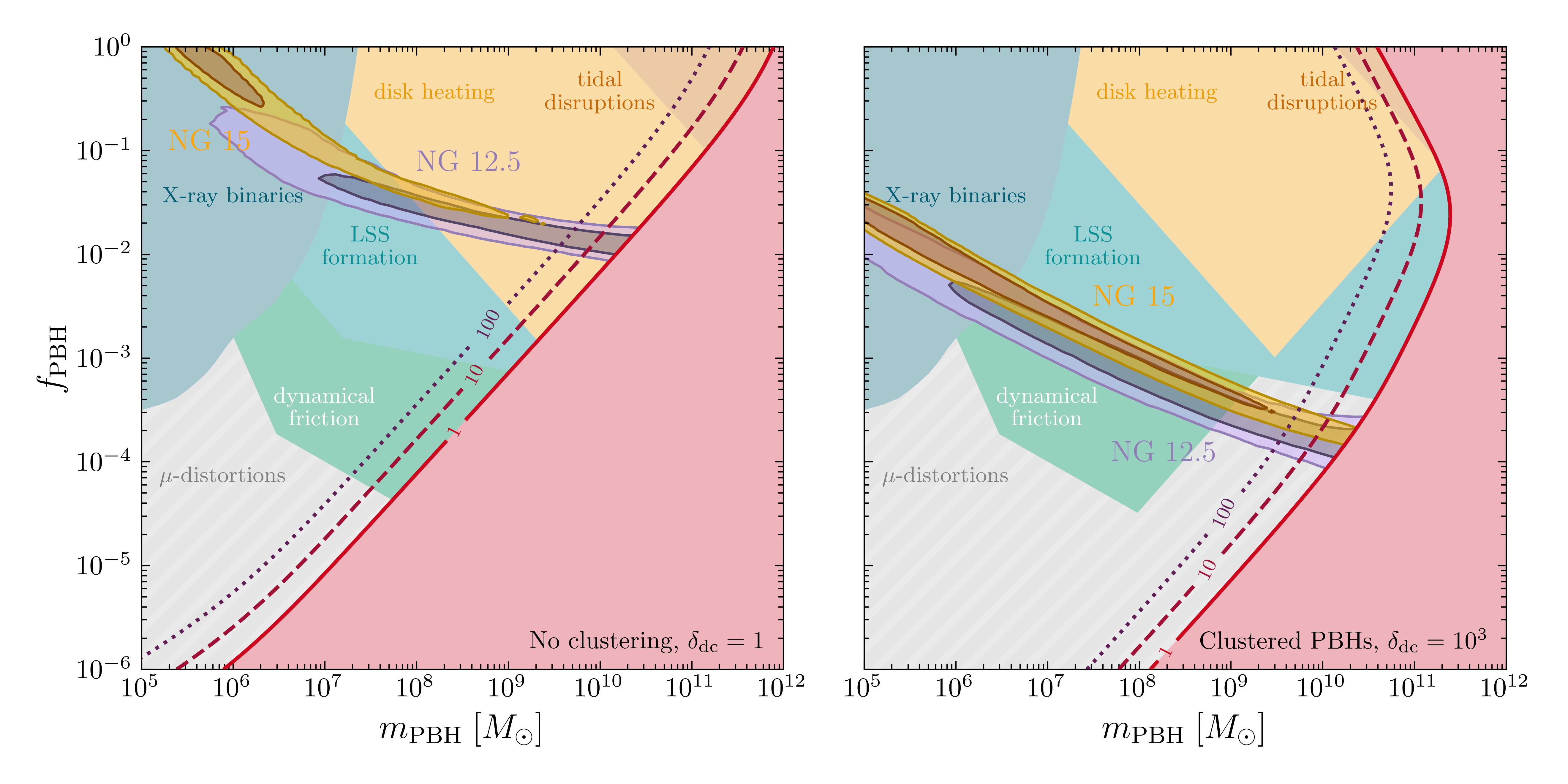}
    \caption{Best-fit region where the NANOGrav 15\,yr (brown) and 12.5\,yr (purple) data set can be explained by merging PBH binaries without clustering (left) and with significant clustering (right). We indicate regions, where $\bar{N} < 1$, $10$, and $100$ PBH binaries are expected to contribute to the signal, and show complementary constraints as discussed in the main text.}
    \label{fig:res_explanation}
\end{figure*}

The most relevant limits in the mass range of interest come from the heating of stars in the Galactic disk~\cite{Carr:1997cn}, the tidal disruption of galaxies~\cite{Carr:1997cn}, the dynamical friction effect on halo objects~\cite{Carr:1997cn}, requiring successful formation of the observed large scale structure (LSS)~\cite{Carr:2018rid}, and observations of x-ray binaries~\cite{Inoue:2017csr}. Many of these limits require at least one PBH per relevant cosmic structure. In the case of PBH clustering, we expect that some structures will contain more than one PBH while others contain no PBH at all. This may shift limits to smaller masses roughly by the ratio of individual PBH and cluster masses. A proper evaluation necessitates further assumptions and detailed simulations, which would also induce dependence on the clustering model. We therefore only show the constraints cited above from simplified analytical treatments without clustering as estimates.

Depending on the production mechanism of the PBHs, strong limits may also arise from the observation of the cosmic microwave background (CMB). In particular if PBHs form due to the tail of Gaussian density fluctuations, Silk damping leads to $\mu$-distortions and strong constraints over a sizable mass range~\cite{Carr:1993aq}. However, this limit crucially depends on the production mechanism and may therefore even be completely evaded. In fact, as discussed above, large clustering generally requires a different production mechanism, calling into question the relevance of these limits.

Additional constraints from future GW observations could lead to further constraints relevant for the PBH parameter space, see the right panel of Fig.~\ref{fig:rate_ogw}. We check explicitly that the Laser Interferometer Space Antenna (LISA) will not be able to probe our best-fit point, but will rather allow independent tests of the parameter space constraints from x-ray binaries~\cite{Inoue:2017csr}.

\paragraph*{PTA data analysis.---}%
We fit the GWB spectrum from PBH binary mergers to the NANOGrav 15\,yr and 12.5\,yr data sets via the interface \texttt{PTArcade}~\cite{mitridate2023ptarcade} for \texttt{CEFFYL}~\cite{Lamb2023} using the 14 (five) lowest Fourier modes for the 15\,yr (12.5\,yr) data set. Given that evidence for a Hellings-Downs correlation is only present in the new data set, we assume a common uncorrelated red noise spectrum for the 12.5\,yr data set and only move to Hellings-Downs correlations for the 15\,yr data set. We validate our approach by comparison to results obtained using \texttt{ENTERPRISE} and \texttt{ENTERPRISE-EXTENSIONS}~\cite{enterprise,enterprise2}. We perform calculations with no clustering, $\delta_\dc = 1$, and significant clustering, $\delta_\dc = 10^3$, choosing log priors $f_\text{PBH} \in [10^{-5},1]$ and  $m_\text{PBH} \in [10^{5},10^{12}] \, M_\odot$. When deriving constraints we further add a power-law signal corresponding to SMBHBs. After marginalizing over the amplitude, the region in the remaining $(m_\text{PBH}, f_\text{PBH})$ plane excluded with $2\, \sigma$ corresponds to the excluded parameter space in which the expected GWB signal from PBH binaries would exceed observations.

\paragraph*{Results.---}%

In Fig.~\ref{fig:res_explanation} we show the regions in PBH mass $m_\pbh$ and DM fraction $f_\pbh$, where the NANOGrav signal can be explained assuming merging PBHs with no clustering (left) and significant clustering with $\delta_\dc = 10^3$ (right). These contours are only expected to be reliable when the expected number of binaries is $\bar{N} \gtrsim 10$ (cf.\ dashed dark red lines) as otherwise the signal observable in NANOGrav is expected to have significant deviations from the global average GWB signal used in our analysis. Particularly, for $\bar{N} < 1$ (full red line) one would not even expect any GWB signal in most realizations of PBH binary distributions. These effects were neglected in~\cite{Atal:2020yic} where it was incorrectly concluded that the NANOGrav signal could be explained in a region of parameter space without clustering, where $\bar{N} \lesssim 1$. We find that the case without clustering cannot explain the NANOGrav signal once taking into account cosmological and astrophysical constraints.

Including clustering shifts the signal regions to smaller $f_\pbh$, enabling a consistent explanation of the PTA data without violating observational constraints, provided that the PBH formation mechanism does not result in significant $\mu$ distortions. As demonstrated in~\cite{Hooper:2023nnl}, certain formation mechanisms for supermassive PBHs can indeed avoid $\mu$-distortion bounds. Note that clustering is also expected to further open up parameter space for a consistent explanation as complementary constraints are expected to shift to smaller masses and the LSS constraints quickly lose sensitivity for masses above $\sim 4 \times 10^{10} M_\odot$~\cite{Carr:2018rid}. While the exact quantitative impact of clustering on these constraints requires detailed, model-dependent simulations of PBH cluster formation, our analysis indicates that the unconstrained best-fit region identified in our work remains robust.

Comparing the results for the 15\,yr and 12.5\,yr data sets we note that the signal regions shift to larger $f_\pbh$, due to the preference for a larger GWB in the new data set~\cite{NANOGrav:2020bcs,NANOGrav:2023gor}, and there is a slight preference for lower masses, where the earlier emission of GWs leads to an increased slope, cf.\ Fig.~\ref{fig:rate_ogw}, as preferred by the new data~\cite{NANOGrav:2023gor}.

\begin{figure}
    \centering
    \includegraphics[width=\columnwidth]{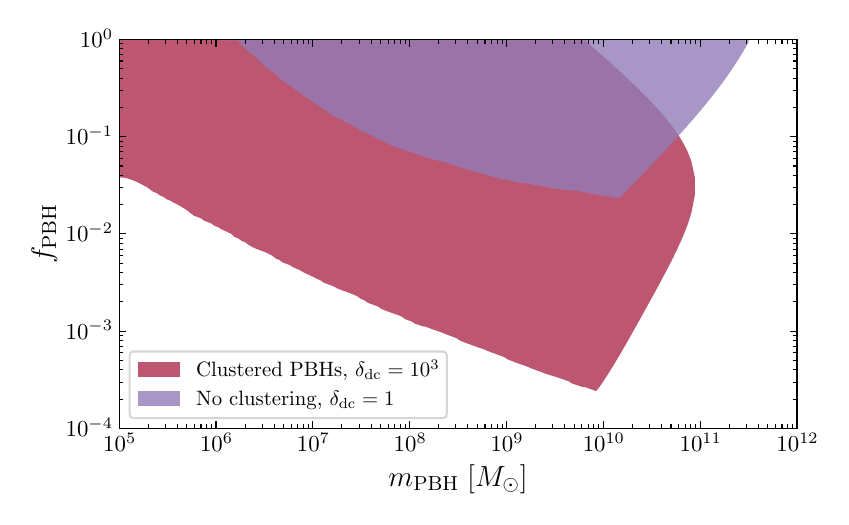}
    \caption{
    Constraints on PBHs from PTA observations with no clustering (purple) and significant clustering $\delta_\dc = 10^3$ (red). We conservatively cut the limits when the expected number of binaries contributing to the signal falls below $10$.}
    \label{fig:constraint}
\end{figure}

If the GWB signal generated by inspiraling PBH binaries becomes too large, above the regions explaining the PTA data in Fig.~\ref{fig:res_explanation}, the abundance of PBHs can be \emph{constrained} by PTA observations. The constraints are shown in Fig.~\ref{fig:constraint} with and without clustering, conservatively requiring $\bar{N} \geq 10$.

\paragraph*{Discussion and conclusions.---}%
In this work we studied the possibility that the signal observed by various PTAs was due to inspiraling primordial supermassive black hole binaries. If the PBHs were ``homogeneously'' distributed at their formation, i.e., followed a Poisson distribution, significant cosmological and astrophysical constraints excluded the possibility of explaining the PTA signal with inspiraling PBHs. Instead considering a clustered spatial distribution of PBHs increased the binary merger rate and thus enabled a consistent explanation of the PTA signals with inspiraling PBH binaries. Crucially, we checked that the signal prediction is also reliable in the relevant parameter space by computing the expected number of binaries contributing to the gravitational wave signal. Further, we used PTA data to constrain the PBH parameter space when the GWB generated during the mergers resulted in stronger signal strengths than the one detected.

This work may serve as a motivation for model builders to construct scenarios that can generate clustered supermassive PBHs without running into cosmological and astrophysical constraints, in particular, due to $\mu$ distortions of the CMB arising from PBH formation. Moreover, our analysis may also motivate future studies to refine astrophysical constraints on clustered PBHs through detailed simulations, as the most stringent existing bounds typically neglect clustering effects. While the latest PTA data finds no evidence for individual compact binary merger events on top of a stochastic background or anisotropies of the GW spectrum, the situation might change in the future. Such a detection would likely invalidate most other cosmological explanations, but is a prediction of our scenario.

\begin{acknowledgments}
\mbox{ }\\
\vfill
 \paragraph*{Acknowledgments.---}%
We would like to thank Andrea Mitridate, Xiao Xue, Joachim Kopp, and Felix Kahlh\"ofer for helpful discussions on related works. This work is funded by the Deutsche Forschungsgemeinschaft (DFG) through Germany's Excellence Strategy EXC 2121 ``Quantum Universe''  Grant No.~390833306 and EXC 2118/1 ``Precision Physics, Fundamental Interactions, and Structure of Matter'' Grant No.~39083149. We acknowledge the use of \texttt{PTMCMC}~\cite{justin_ellis_2017_1037579} for sampling and \texttt{NUMPY}~\cite{harris2020array}, \texttt{MATPLOTLIB}~\cite{Hunter:2007}, and a customized version of \texttt{ChainConsumer}~\cite{Hinton2016} to analyze and visualize our results.
\end{acknowledgments}

\appendix

\section{Appendix: Number of contributing PBH}
\label{app:N}

Here we derive the expression for the expected number of primordial black hole binaries contributing to the gravitational wave background within a given frequency range. To do so, recall that $R(t_\text{r} + \tau_{f_\text{r}})$ is the number of binaries per comoving volume $V_\text{c}$ and per time interval in the cosmic rest frame $t_\text{r}$ emitting with a frequency $f_\text{r}$, where $\tau_{f_\text{r}}$ is the time until coalescence. Hence, the expected number of binaries $\d \bar{N}$ emitting at redshifts between $z$ and $z + \d z$ with a frequency within the logarithmic interval $\d \ln f_\text{r}$ around $f_\text{r}$ is~\cite{Sesana:2008mz}
\begin{align}
    R (t_\text{r} + \tau_{f_\text{r}}) &= \frac{\d z}{\d t_\text{r}} \frac{\d^2 \bar{N}}{\d z \, \d \ln f_\text{r}} \frac{\d \ln f_\text{r}}{\d t_\text{r}} \frac{\d t_\text{r}}{\d z} \frac{\d z}{\d V_\text{c}} \nonumber \\
    &= \frac{\d^2 \bar{N}}{\d z \, \d \ln f_\text{r}} \left( - \frac{\d \ln f_\text{r}}{\d \tau_{f_\text{r}}} \right) \frac{\d z}{\d V_\text{c}}\eqsp,
\end{align}
where we used that the time in the cosmic rest frame a source is emitting within the frequency interval is given by
\begin{align}
    \d t_\text{r} \frac{\d \ln f_\text{r}}{\d t_\text{r}} = - \d t_\text{r} \frac{\d \ln f_\text{r}}{\d \tau_{f_\text{r}}}\eqsp.
\end{align}
Since the binaries emitting in the logarithmic frequency interval $\d \ln f_\text{r}$ around $f_\text{r}$ are detected today in a logarithmic frequency interval $\d \ln f$ around $f$, where $\d \ln f_\text{r} = \d f_\text{r} / f_\text{r} = \d f / f = \d \ln f$, we have
\begin{align}
    \frac{\d^2 \bar{N}}{\d z \, \d \ln f} = \frac{\d^2 \bar{N}}{\d z \, \d \ln f_\text{r}} \;.
\end{align}
With Eq.~\eqref{eq:tau_f_r} and the definition of the comoving volume in a spatially flat universe
\begin{align}
    V_\text{c} (z) = \frac{4 \pi}{3} [d_\text{c}(z)]^3 = \frac{4 \pi}{3} \left( \int_0^z \frac{\d z'}{H(z')} \right)^3 \eqsp ,
\end{align}
where $d_\text{c}$ is the comoving distance and $H$ is the Hubble rate, we find
\begin{align}
    \frac{\d^2 \bar{N}}{\d z \, \d \ln f} = \frac{8}{3} \, \tau_{f_\text{r}} \, \frac{4 \pi \, [d_\text{c}(z)]^2}{H(z)} \, R (t_\text{r} - \tau_{f_\text{r}})\eqsp.
\end{align}
The average number of binaries $\bar{N} (f_-, f_+)$ contributing to GWs of frequencies between $f_-$ and $f_+$ is therefore given by
\begin{align}
    \bar{N} (f_-, f_+) = \! \int_{f_-}^{f_+} \frac{\d f}{f} \int_0^\infty \d z \, \frac{32 \pi \tau_{f_\text{r}} \, [d_\text{c}(z)]^2}{3 H(z)} R (t_\text{r}- \tau_{f_\text{r}})\eqsp.
\end{align}
Note that the \textit{actual} number of binaries is then Poisson distributed with mean $\bar{N} (f_-, f_+)$.

\newpage

\bibliography{bibliography}

\end{document}